# CHARACTERISATION OF SURFACE LAYERS FORMED ON PLASMA-FACING COMPONENTS IN CONTROLLED FUSION DEVICES: ROLE OF HEAVY ION ELASTIC RECOIL DETECTION

***P. Ström[1], P. Petersson[1], M. Rubel[1], A. Weckmann[1],***

***S. Brezinsek[2], A. Kreter[2], S. Möller[2], K. Rozniatowski [3]***

*[1]Department of Fusion Plasma Physics, Royal Institute of Technology (KTH),*

*Teknikringen 31, 100 44 Stockholm, Sweden*

*[2]Forschungszentrum Jülich GmbH, Institut für Energie- und Klimaforschung –*

*Plasmaphysik, 52425 Jülich, Germany*

*[3]Faculty of Materials Science, Warsaw University of Technology,*

*PL-02-507 Warsaw, Poland*

Corresponding author

Petter Ström, pestro@kth.se

Tel. +468 790 6111

**Abstract**

Wall components retrieved from the TEXTOR tokamak after tracer experiments with nitrogen-15 and molybdenum hexafluoride ($MoF_6$) injection were studied to determine deposition patterns and, by this, to conclude on material migration. Toroidal limiter tiles made of carbon fibre composites and fine grain graphite were examined using time-of-flight heavy ion elastic recoil detection analysis. Molybdenum deposition patterns indicated migration based on erosion and prompt re-deposition. Nitrogen-15 was trapped together with the deposited molybdenum. Some information on the depth distribution of species in the top 400 nm layer of the limiters was obtained; however surface roughness of the samples strongly limited resolution. In the case of molybdenum, the largest concentration was found in the 100 nm outermost layer, whereas fluorine and nitrogen-15 displayed more irregular profiles. Other species, besides deuterium fuel and carbon-12, were also identified: boron-10 and boron-11 originating from boronisations, carbon-13 from earlier tracer experiments, nitrogen-14 from plasma edge cooling and metals eroded from the Inconel wall.

## 1. Introduction

Controlled fusion devices with magnetic confinement feature toroidal vacuum vessels that belong to the largest vacuum systems used in science and technology today. For example, the vessel volume is 100 $m^3$ in the Joint European Torus (JET) which is the largest present-day tokamak, while around 1600 $m^3$ is the design value for the International Experimental Thermonuclear Reactor (ITER) [1]. Consequently, the surface area of the plasma-facing wall is of the order of tens of square meters in smaller fusion devices, nearly 200 $m^2$ in JET and 850 $m^2$ is projected for ITER. The structure of the wall is highly complex. This is demonstrated in Fig. 1 a, which shows a toroidal view into the vacuum vessel of TEXTOR, a medium-sized tokamak (volume 10 $m^3$, wall area 37.5 $m^2$) that was operated until December 2013 at Forschungszentrum Jülich, Germany [2].

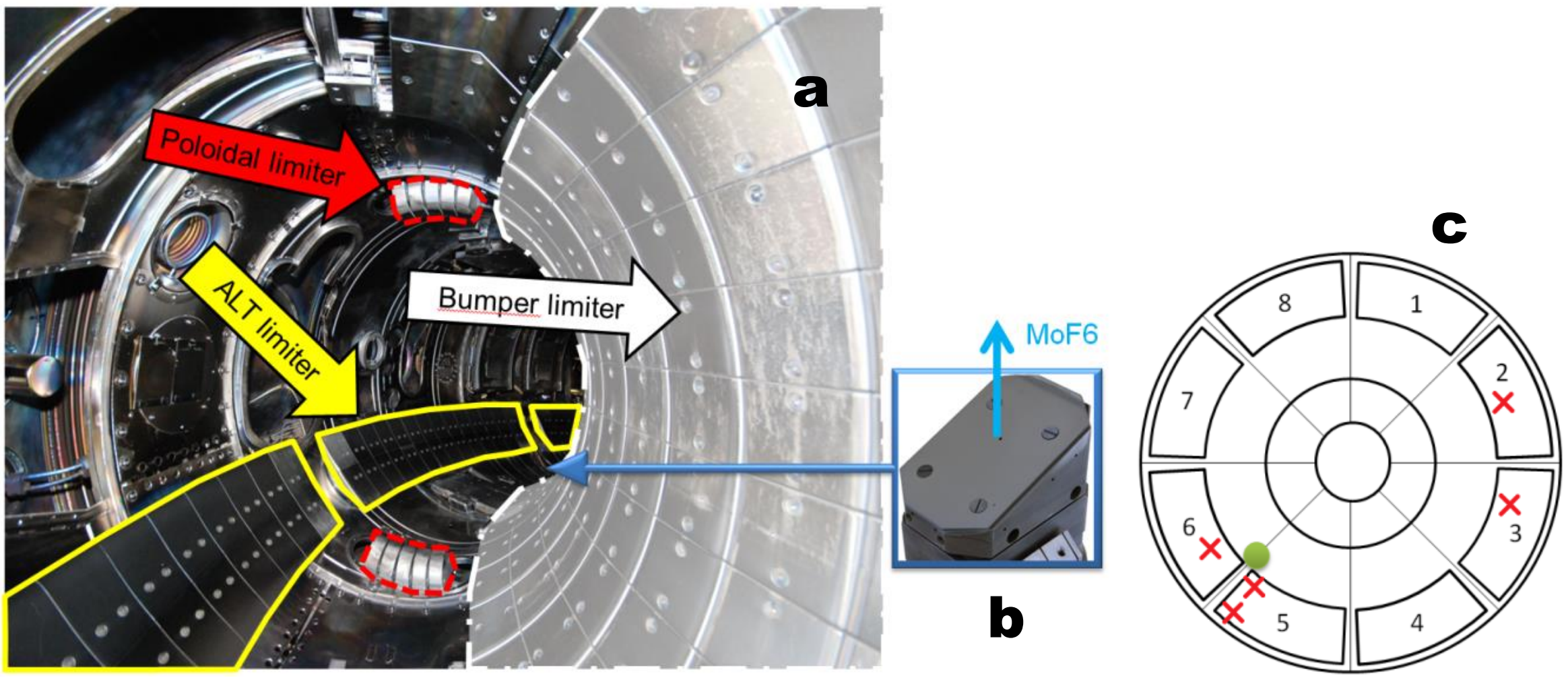

Figure 1. (a) Toroidal view inside the TEXTOR vacuum vessel with all limiters (including the test limiter) marked, (b) a magnified image of the $MoF_6$ inlet and (c) a scheme of the ALT-II limiter seen from above, with the location of the $MoF_6$ gas inlet marked by a small circle and the analysed tiles marked with crosses.

The major plasma-facing components (PFCs) are the limiters of various kinds (marked in Fig. 1 a), i.e. high heat flux (HHF) components that *limit* the plasma to protect other components from the direct impact of particle fluxes. As a result of plasma-wall interaction (PWI), plasma-facing components, especially HHF parts, undergo significant modification [3-7] under the impact of energetic particles. Material erosion occurs, followed by the transport of eroded species (ionized atoms and molecules) in the torus and, eventually, the deposition of those species that were not pumped out. Such deposition often occurs together with fuel atoms, i.e. hydrogen isotopes, forming mixed material layers referred to as co-deposits. The properties of co-deposits usually differ significantly from those of the original wall material. Considering these facts, topics of primary interest in PWI studies include the assessment of fuel retention in the wall and material lifetime. The latter calls for a detailed determination of the entire material migration cycle: erosion – transport –deposition – re-erosion, etc. [3,4,8,9]. These processes occur for every material exposed to plasma, but to various extents [4]. There is no ideal wall material that can meet a full range of requirements including high thermal conductivity, resistance to thermal shocks, low radiation losses by eroded impurity species, low erosion rate and low fuel retention. Carbon materials, especially graphite and carbon fibre composites, were considered as primary choices for a long time. As such, they were installed in most tokamaks, also in TEXTOR and, until 2009, in JET. However, high fuel retention connected to the formation of hydrocarbon layers motivated tests of other low-Z candidates, especially beryllium for areas of moderate power loads (below 5 MW $m^{-2}$) and tungsten for HHF components. These two materials have been used since 2011 for the ITER-Like Wall in JET [10,11] and, after successful experiments [12,13],  the same combination has been chosen for the wall of ITER.

Preparations for ITER and subsequent reactor operation require comprehensive research regarding the behaviour of high-Z metals in tokamaks, e.g. fuel retention and material migration under the impact of plasma impurities such as carbon, oxygen or other gases injected deliberately into the torus for plasma edge cooling [6,7,9,14]. The last category comprises predominantly nitrogen but also neon and heavier noble gases are tested. As a consequence, questions appear regarding nitrogen transport, reactivity with wall components and deposition that may lead to long-term retention. To answer these questions, tracer techniques are used [14,15]. They are based on the design of dedicated experiments in which one applies various marker tiles, wall probes and/or marker gases. A normal course of action is to develop such methods in a medium-size machine and then transfer the procedures to a large tokamak. The techniques have, thus, been developed and tested in TEXTOR, whose mission was strongly oriented towards PWI studies. Over the years, many experiments have been performed with rare, non-radioactive isotopes such as carbon-13 in the form of $^{13}CH_4$ to determine carbon transport [6-8,16-19], nitrogen-15 [20-24] and oxygen-18 [2,24,25]. For tungsten transport studies, a volatile hexafluoride, $WF_6$, has been used [9,26].

Following an experimental campaign, carried out with or without markers, some in-vessel components are retrieved from the torus and studied ex-situ. In the design of marker experiments there are at least two indispensable elements: availability of a marker and availability of methods to detect and measure it quantitatively on surfaces. Accelerator-based ion beam analysis (IBA) methods are very useful in studies of PFCs, as described earlier [27-29]. They are quantitative, highly sensitive, allow depth profiling even up to over 10 micrometers in some substrates and one can selectively determine light isotopes on surfaces. For instance, the analysis of deuterium by nuclear reactions is here of extreme importance. In the case of tracer experiments performed at the very end of experimental campaigns, one is interested in the composition of the surface region of PFCs or special probes. In this field, time-of-flight heavy ion elastic recoil detection analysis (ToF-HIERDA) plays special role, as in a single measurement one may quantify light isotopes with atomic mass up to about 20 amu and also detect heavy species, although with poorer mass resolution (see Figures 3-4 and the related paragraphs below for a more in-depth discussion on resolution). A detailed description of the technique is presented in [30].

The aim of this work was to determine, by ToF-HIERDA, the atomic composition of PFCs from TEXTOR in the 400 nm layer closest to the surface, and on this basis to conclude on high-Z material migration as well as to verify the particular usefulness of the technique in detecting light tracers. The distributions of a tracer, nitrogen-15, and a high-Z element, molybdenum, have therefore been in focus. Other species: boron, used for wall conditioning [31,32], $^{14}$N from edge cooling discharges and fluorine from the molybdenum hexafluoride that was used to provide molybdenum in a gas state, have also been studied.

## 2. Experiment

The experiments were performed in the TEXTOR tokamak (see Fig. 1 a). Its main PFCs are: the belt toroidal limiter, also known as the Advanced Limiter Test (ALT-II), poloidal limiters at the top and the bottom of the machine and the inner bumper limiter being a shield for the Dynamic Ergodic Divertor (DED). Molybdenum hexafluoride and nitrogen-15 were puffed into discharges heated by neutral beam injection. The procedure was similar to that employed in the case of $WF_6$ puffing in earlier experiments [9]. Tungsten has been used in TEXTOR in many other experiments and, as a consequence, a relatively high W background was expected to be seen on many PFCs. This fact motivated the choice of molybdenum to examine high-Z transport as a substitute for tungsten.

$MoF_6$ and $^{15}N_2$ were introduced during 31 and 22 discharges respectively. The typical discharge time was 8 seconds. $^{15}N_2$ puffing took place between 1 and 5 seconds, whereas $MoF_6$ injection was done between 0.8 and 1.8 seconds (in all but two discharges, during which it lasted until 2.8 seconds instead). A total of $1.4 \cdot 10^{21}$ $MoF_6$ molecules and $5.3 \cdot 10^{21}$ $^{15}N$ atoms were estimated to have been introduced into the machine. Approximately 30% – 50% of the $MoF_6$ entered the plasma and the rest is assumed to have been retained in the inlet system.

$MoF_6$ was introduced from a localized source using a gas inlet in a roof-shaped test limiter (shown in Fig. 1 b), operated from a limiter lock located in the bottom of the machine: between Blade 5 and Blade 6 of ALT-II, as marked in Fig. 1 c. There are two rows of tiles on each such blade, the upper being numbered 1 to 14 from left to right (clockwise) and the lower 15 to 28. In the following, tiles are referenced by a number combination X-Y, where X is the ALT-II blade number and Y is the tile number on the blade. $^{15}N$ was inserted via Gas Feed 3 (Gaseinlass 3: GE3), located in the equatorial plane, 90$^{o}$ toroidally 'upstream' wrt. the plasma current direction from the $MoF_6$ inlet.

The experiment took place on the last operation day of TEXTOR, before final shutdown. This situation has created a unique opportunity to retrieve a large number of components for ex-situ studies. Five ALT-II tiles from various poloidal and toroidal locations were considered for the studies here described; they are marked with crosses in Fig. 1 c. They were taken from the very vicinity of the $MoF_6$ inlet (corner Tiles 5-14 and 5-28), Tile 19 from the center part of the bottom row on Blade 6 and from toroidally distant locations on Blades 2 and 3. Table 1 contains the tile numbers as well as short descriptions of the motivation for selecting them.

| **Tile<br>(ALT-II blade number – tile number)** | **Description** |
|---|---|
| 5-28 | Closest tile to $MoF_6$ injection point, visible deposited layers. |
| 5-14 | Close to $MoF_6$ injection point, visible deposited layers. |
| 6-19 | Large molybdenum deposits seen using RBS. |
| 3-19 | Far from $MoF_6$ injection point, but clearly visible deposits. |
| 2-21 | Far from $MoF_6$ injection point, no visible deposits (reference tile). |

Table 1. List of tiles selected for detailed analyses and justification of the selection.

Several techniques, which are not further treated here, have been used to study various machine components. These include Rutherford backscattering spectroscopy (RBS), nuclear reaction analysis (NRA) and electron beam methods. As aforementioned, this work was concentrated on the distribution of species in the very surface layer (top 400 nm). Therefore, studies using ToF-HIERDA have been performed with a 36 MeV iodine beam, $^{127}I^{8+}$. Four samples, labeled A to D, were cut out of each tile, as shown in Fig. 2, and used for the analysis.

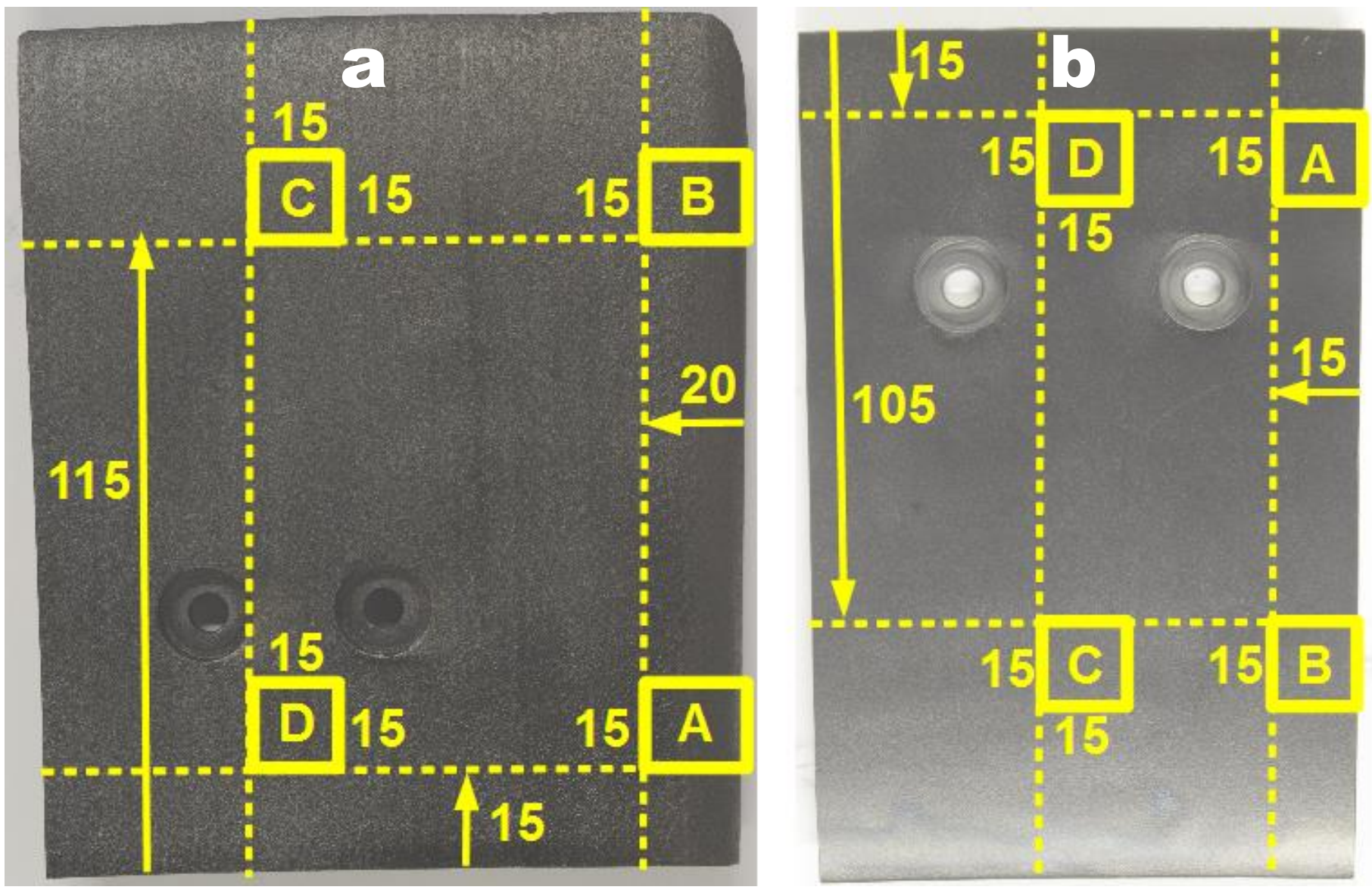

Figure 2. Sample positions on ALT tiles: (a) Blade 5, Tile 14: upper corner tile,

(b) Other tiles (including lower corner Tile 5-28, where same scheme was applied).

Directions are such that the right edges of tiles here displayed are pointing in the clockwise direction in Fig. 1c.

The horizontal position of Samples A and B is at the right edge of the tile, whereas Samples C and D are located on a vertical line running midway between the two bolt holes. Furthermore, Samples A and D originate from the region where deposition has usually been measured, while B and C are in the region of erosion, i.e. the area of direct plasma impact [5,9,33] (these are referred to simply as “erosion zone” and “deposition zone” below). There are two kinds of ALT tiles: corner tiles made of carbon fibre composites (CFC), four at each blade, as exemplified by Tile 5-14 in Fig. 2 (a) and standard tiles made of graphite shown in Fig. 2 (b). The beam was hitting the samples approximately in the middle, with an estimated uncertainty in the position of up to 5 mm, at an angle of 22.5 degrees wrt. the surface, and the detection system was placed at 45 degrees from the primary beam direction, in the plane defined by the beam and sample normal. When discussing ERDA results, the surface roughness of the studied samples must be addressed. The reason is simply that at a shallow angle of incidence, such as in this case, a rather small imperfection in the surface may have a large effect on the amount of material that the beam passes through before interaction at a given depth. This impacts severely the depth resolution of the method. In addition to the ERDA measurement, therefore, surface roughness was recorded for six samples, including both CFC and graphite in the erosion and deposition zones, with a Wyko 9300 optical profilometer from Veeco Instruments.

## 3. Results and discussion

An example of a ToF-HIERDA spectrum, recorded for Sample A of Tile 5-14, is shown in Fig. 3. The excellent resolution that can be obtained for light elements is apparent as counts corresponding to isotopes with mass differences of 1 amu are separated at least for elements with atomic numbers up to 7 (nitrogen). Every element found on the tiles has been marked in the spectrum. Molybdenum and nitrogen-15, which are of particular interest for this study, are highlighted by underlined, red font.

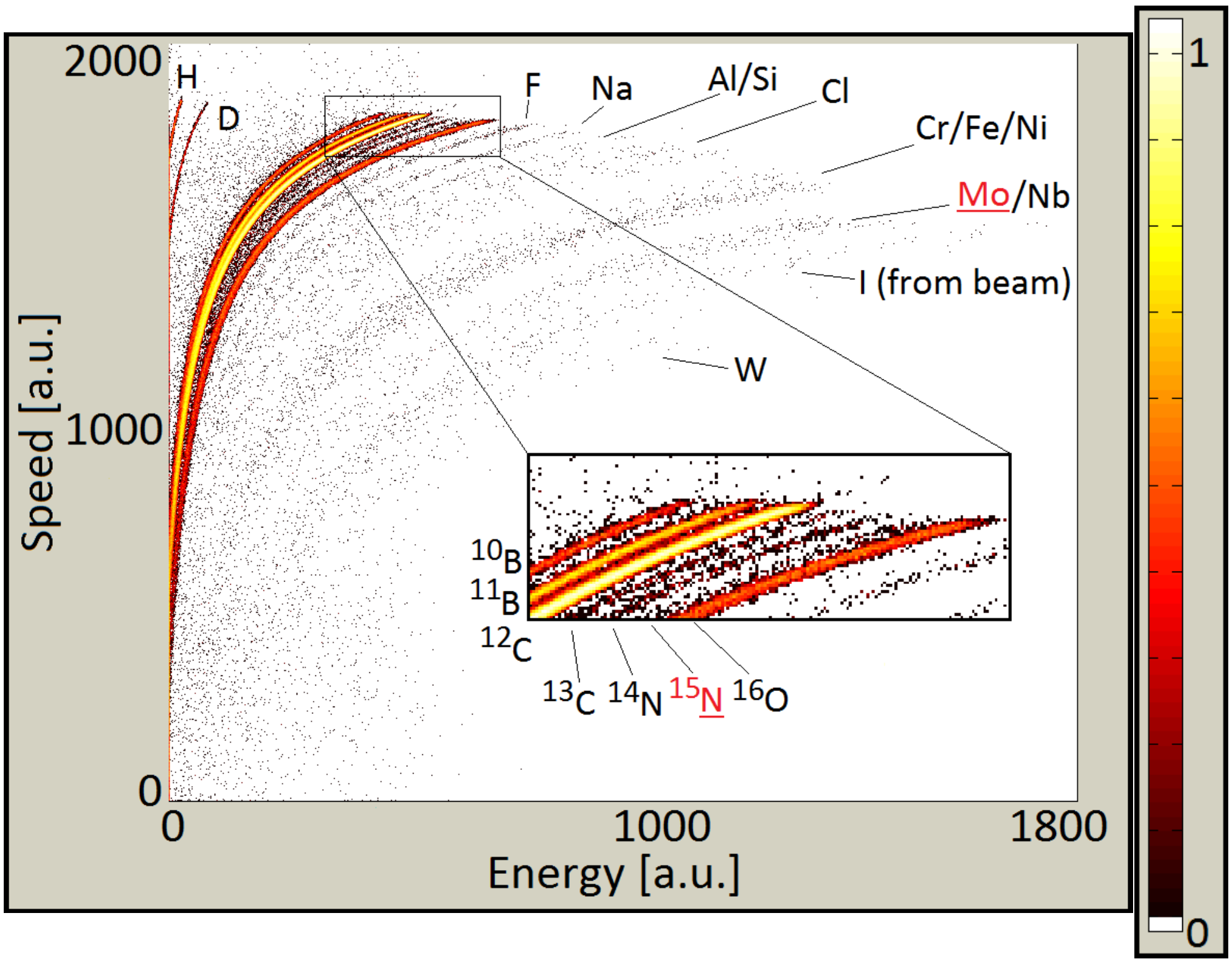

Figure 3. ToF-HIERDA spectrum from Tile 5-14, Sample A. Intensity scale: number of counts in channel relative to maximum number of counts.

Fig. 4 displays zoomed in views of the region from $^{12}$C to $^{16}$O, very close to the surface (approximately top 200nm) of samples 3-19C and 5-28A. The overlaid boundaries enclose counts that come from $^{14}$N (upper area to the left, blue line) and $^{15}$N (lower area to the right, red line). Sample 3-19B features a relatively low amount of $^{15}$N, whereas 5-28A has the highest concentration out of all studied samples (see Fig. 8b).

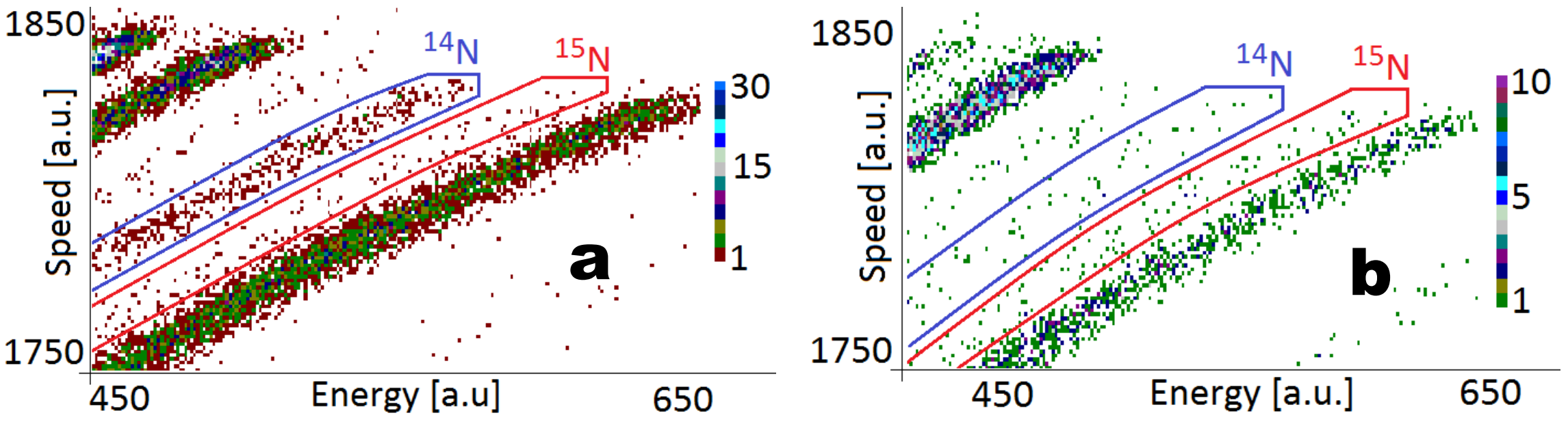

Figure 4. Counts corresponding to $^{14}$N and $^{15}$N for (a) Sample 3-19C, with a moderate amount of $^{15}$N and (b) Sample 5-28A, where $^{15}$N concentration is comparable to that of $^{14}$N.

A comparison between Figures 4 a and b reflects the criterion for data acquisition in this case. Counts were collected until a predefined region of interest corresponding to $^{15}N$ had 200 counts (i.e. a statistical error of 7.1% in the total concentration). As a result, a high concentration of $^{15}N$ is not seen as a large number of counts for that isotope here, but rather a smaller number of counts for everything else. To be clear: the irradiation time and thereby the total number of ions hitting the surface was different for different samples. It was selected as a compromise between getting a low statistical error for $^{15}N$ (which requires a long irradiation time) and minimizing surface modification of the samples during the measurement (which requires a short irradiation time). It should also be noted that the error caused by counts belonging to $^{16}O$ that are counted as $^{15}N$ is significant, especially for large depths. At the very surface, the separation between $^{15}N$ and $^{16}O$ is adequate for the correct classification of the majority of counts whereas beneath 400nm, the two tracks start to merge considerably.

Addressing the question of surface roughness, 3D maps of a graphite tile (6.19) and a CFC tile (5.14) are shown in Figures 5 a and b. In both cases, the scanned area is 0.9 x 1.3 mm$^2$, i.e. approximately the size of the beam spot and the height scale has been magnified so that micrometer sized structures can be seen. The largest roughness, especially on CFC tiles, is found in pits on the surface whose depths are in the order of a few tens of micrometers. A large portion of these pits is inaccessible to the ERDA measurements here performed. Information comes rather from the flatter regions between the pits. Even in those regions, however, there are height variations in the order of several micrometers. The resulting inclination of the surface gives rise to an illusion of depth, since either the beam or the outgoing particles have to pass through some material even when an interaction takes place right at the surface.

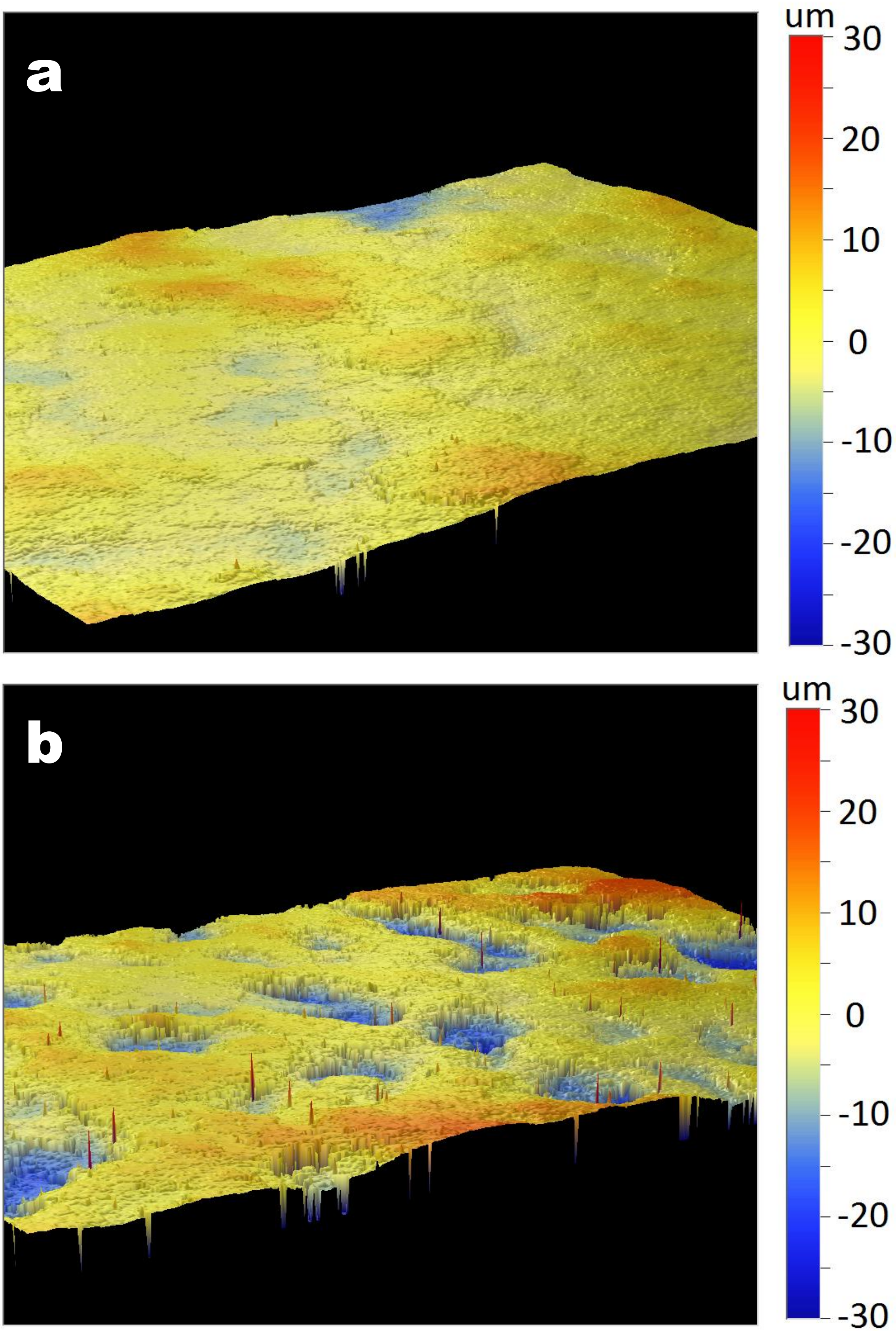

Figure 5. Surface profiles on (a) Sample 6-19A (graphite), and (b) Sample 5-14D (CFC). Obtained with a Wyko 9300 optical profilometer. Note that the thin spikes visible especially in (b) are artefacts from the measurement.

In table 2, the average deviation from the mean surface ($R_a$) is given, for three scans on different positions on six samples: three from the erosion zone and three from the deposition zone on the CFC tile 5-14, the graphite tile 6-19, both with relatively large amounts of deposits on them, and the cleaner graphite tile 2-21.

| **Sample** | **Description** | **$R_a$, scan 1 [µm]** | **$R_a$, scan 2 [µm]** | **$R_a$, scan 3 [µm]** |
|---|---|---|---|---|
| 5-14C | CFC tile, erosion zone | 8.31 | 8.40 | 10.50 |
| 5-14D | CFC tile, deposition zone | 5.41 | 6.21 | 6.49 |
| 6-19A | Graphite tile, deposition zone | 3.21 | 3.49 | 3.30 |
| 6-19C | Graphite tile, erosion zone | 1.91 | 2.22 | 2.24 |
| 2-21C | Graphite tile, erosion zone | 2.48 | 2.35 | 2.10 |
| 2-21D | Graphite tile, deposition zone | 1.81 | 1.87 | 1.80 |

Table 2. Surface roughness on six samples quantified by average deviation from the mean surface.

As a result of the high irregularity of the sample surfaces, we conclude that depth profile measurements in this case will have a very bad resolution. Also, the error in the absolute amount of any species per unit area is hard to assess. The relative amounts of different species, however, are accurate, as we show with the example of boron isotopes below. Before any detailed analysis, we return now to explaining the occurrence of elements as seen in Fig. 3.

Hydrogen and deuterium are commonplace in TEXTOR plasmas. Boron is present due to boronisations [31,32] that have been performed regularly using a diborane–helium gas mixture (usually 10% $B_2H_6$ and 90% He), as a method of wall conditioning between 5 and 10 times per year. The carbon that constitutes the limiter tiles themselves is clearly visible in Fig. 3. The feature of $^{13}C$ is associated both with the occurrence of that isotope in carbon (natural abundance 1.109 %) and earlier experiments on transport studies with an isotope-labeled $^{13}CH_4$ [34,35]. Also the detection of $^{14}N$ is most probably related to two sources of that isotope: (a) adsorption of nitrogen from air when the tiles were exposed to ambient atmosphere after the retrieval from the torus; (b) co-deposition of species retained in the TEXTOR wall during earlier discharges with $^{14}N_2$ puffing to cool the plasma edge; that effect has been reported previously [9,36]. Even isotope exchange with air nitrogen cannot be excluded but its extent is not assessed here. In order to quantify nitrogen retention and to unambiguously distinguish between nitrogen originating from air and that co-deposited during experiments, a tracer technique with $^{15}N_2$ injection is used (The natural abundance of $^{15}N$ is 0.37%). The situation with oxygen is similar, as it is known to be incorporated in co-deposits from air (water vapour and oxidation of species) [37]. Therefore, $^{18}O_2$ (natural abundance 0.205 %) was used earlier in some cleaning discharges [24,25] and traces of that isotope remained in the vessel until the experiment with $^{15}N$ and $MoF_6$ reported here. One finds fluorine ($^{19}F$) from $MoF_6$ and molybdenum itself. The latter comes mainly from the hexafluoride puffing though there is also some Mo background associated with the presence of that metal as an additive in the Inconel-625 alloy that is the primary construction material of the liner. RBS measurements on samples from before the experiment have indicated an average Mo background of ca. 150 ppm, corresponding to $6 \times 10^{14}$ cm$^{-2}$ for the top 400nm layer, and a peak value of 400 ppm, or $1.6 \times 10^{15}$ cm$^{-2}$. The Mo trace in the spectrum is fairly broad because of six natural isotopes and also because of the presence of niobium eroded from Inconel. Since Inconel-625 contains less than half as much niobium as it does molybdenum, we assume that the niobium background is negligible (less than the level of molybdenum detected on tiles 3-19 and 2-21, see Fig. 8 b). Furthermore, earlier chemical analysis with inductively coupled plasma mass spectrometry (ICP-MS) of flaking deposits, scraped from limiter tiles, has shown a niobium to molybdenum ratio of only 15%. Nickel and chromium, as the main constituents of the aforementioned Inconel alloy, are found together with iron in all co-deposits on plasma-facing components of TEXTOR [20,37]. Sodium and chlorine are present in relatively small concentrations, probably originating from intrinsic impurities of the wall components. Aluminum and silicon, whose features cannot be quantitatively resolved, may originate from eroded/damaged ceramic elements in diagnostic and liner baking systems. Silicon could potentially also be present due to past siliconisations [38-40]. Finally there is a small amount of tungsten which was used in previous experiments with various solid tungsten limiters [41-43] and $WF_6$ injection experiments [9]. Notably, this tungsten background is not as severe as was expected beforehand. The presence of iodine in the spectrum is related to forward scattered particles from the analyzing beam; there is no iodine in the studied limiters.

By numerically dividing the top 400 nm of the samples into 15 depth-slices (about 27 nm thick) and taking into account the kinematics and stopping power of ions in the sample matter, an iterative calculation was performed, yielding the atomic concentrations of all identified species in the slices. Depth profiles of all deposited materials have thereby been obtained. The calculations were performed using the CONTES code for MATLAB, originally created by Martin Janson, updated to the here used version 15.2 at Uppsala University as well as the Royal Institute of Technology (KTH) in Stockholm, with stopping power data from SRIM [44], version 2012.01. The results were double checked against CONTES version 07, which uses stopping powers from SRIM 2006. Aluminum was disregarded in this calculation due to an uncertainty about whether the aluminum signal is a false one originating from aluminum parts in the ToF-HIERDA set-up. Also tungsten was ignored because of its very small abundance in conjunction with the difficulty to identify the few spread-out tungsten counts (see, again, Fig. 3). All other species were taken into account. Ignoring Al and W is of no consequence for the results as, due to very low concentrations, these species do not significantly alter stopping power in the samples or the normalization. For the samples we have studied, the maximum information depth is around 800 nm, but we still chose to work only down to 400 nm to keep a fairly good separation of counts for different species.

Two examples of depth profiles for molybdenum, fluorine and nitrogen-15 are shown in Fig. 6 a-b for Tiles 5-28 and 6-19, both close to the gas inlet. The relative atomic contents of the shown species are small because they are not major constituents of the carbon- and boron-rich layers. The depth scale is given in units of $10^{15}$ atoms/cm$^2$; one such unit corresponds roughly to 1 Å of depth. It should be stressed again that surface roughness makes the details of these figures, as well as Fig. 7, unreliable but some features can be noted. In the case of molybdenum, Fig. 6 b is representative for samples on which a considerable amount of this element is found (more than 3 x $10^{15}$ cm$^{-2}$); it is seen primarily in the outermost layer of about 100 nm thickness. For places in the very vicinity of the $MoF_6$ inlet (Tiles 5-14 and 5-28), the maximum of $^{15}N$ occurs before or close to 100 nm (Fig. 6 a), while in more distant locations smaller amounts are trapped and the profile is simply without a clear pattern (Fig. 6 b). Fluorine often also has a maximum near the surface within 100 nm but its depth profiles are not showing a consistent pattern for any group of samples. The possibility that concentrations of these elements actually peak at the surface, but that the distributions are shifted due to surface roughness and apparent depth (as explained above), should not be dismissed. We note here the timeline of plasma operation and the gas feed to the torus, especially during the final stages of the experiment: molybdenum hexafluoride was puffed between Blades 5 and 6 until the last discharge, while the $^{15}N_2$ injection from GE3 was stopped five discharges earlier because of the wall saturation with nitrogen which then required glow discharge cleaning with helium. The earlier termination of $^{15}N$ usage is, however, not discernible in our results.

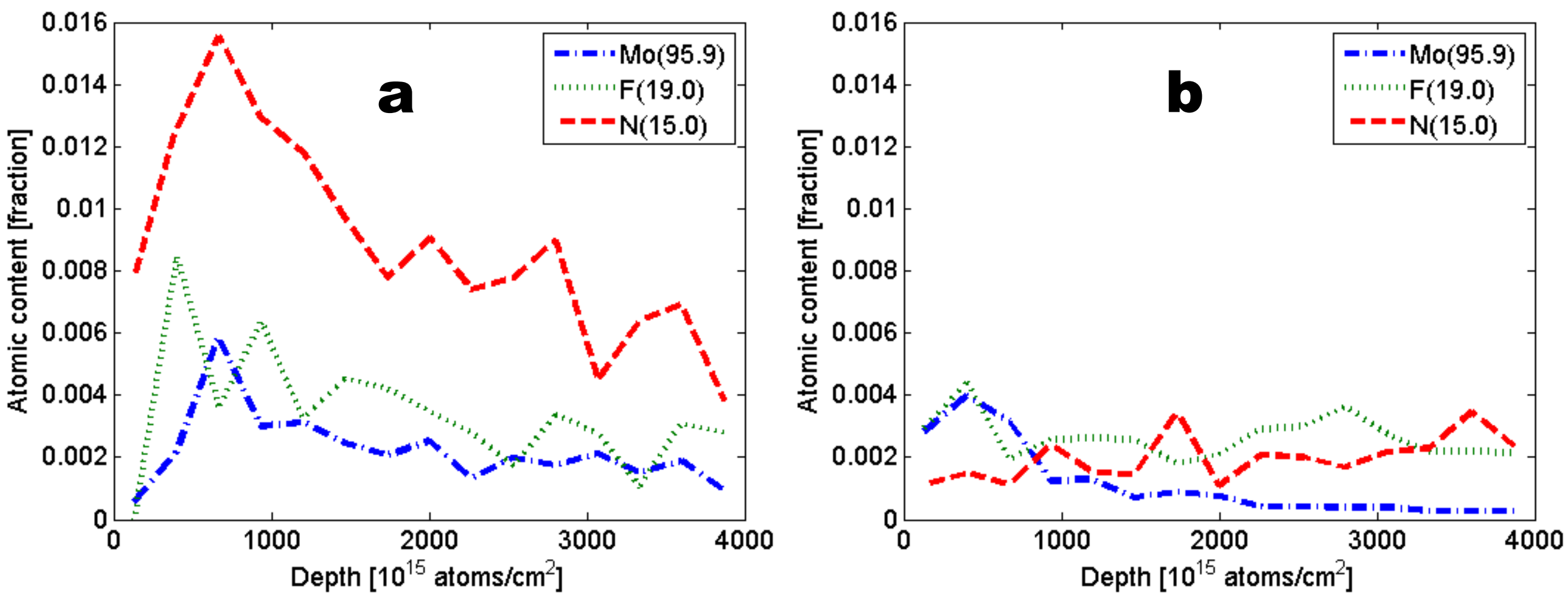

Figure 6. Depth profiles of Mo, $^{19}F$ and $^{15}N$ on (a) Tile 5-28, Sample A and (b) Tile 6-19, Sample D

Fig. 7 shows depth profiles of carbon-12, boron-11, boron-10 and hydrogen, i.e. major species on the tiles. In contrast to the surface peaking of molybdenum, these elements are present all the way through the displayed 400 nanometers, but their relative contents may vary several tens of percent. The plots clearly show that boron maxima are in anticoincidence with carbon (simply because boron claims a considerable fraction of all atoms from the carbon substrate). There is no very clear pattern over all samples when it comes to the carbon and boron structures. As such, Figures 7 a and b should be treated more as examples than representative for a general trend. Especially the surface of Tile 3-19 contains significant amounts of boron isotopes (with more boron than carbon for at least 400 nm). The non-uniform, "wavy" distributions of boron and carbon, seen in Fig. 7 a, were observed in the past in relation to boronisations, when thin boron-rich strata separated carbon rich co-deposits of thicknesses between 2 and 4 micrometers [7]. However, the features measured with ToF-HIERDA are separated by only a fraction of a micrometer and thus cannot be related directly to earlier consecutive boronisations of TEXTOR. They rather reflect the situation after the last such boronisation, performed only 92 discharges before the marker experiment. On Sample 3-19C, from the erosion zone, the boron and carbon profiles could be tentatively attributed to events which have an impact on the chemistry of erosion of various species, e.g. disruptions. However, there is generally no significant difference seen between these profiles on samples from the erosion zone and the deposition zone. On Tile 6-19, Sample D, carbon prevails in the surface up to 100 nm and then it is mixed with boron. Note that not all elements that were included in the analysis are displayed here. As such, the curves showing atomic fractions do not add up to 1.

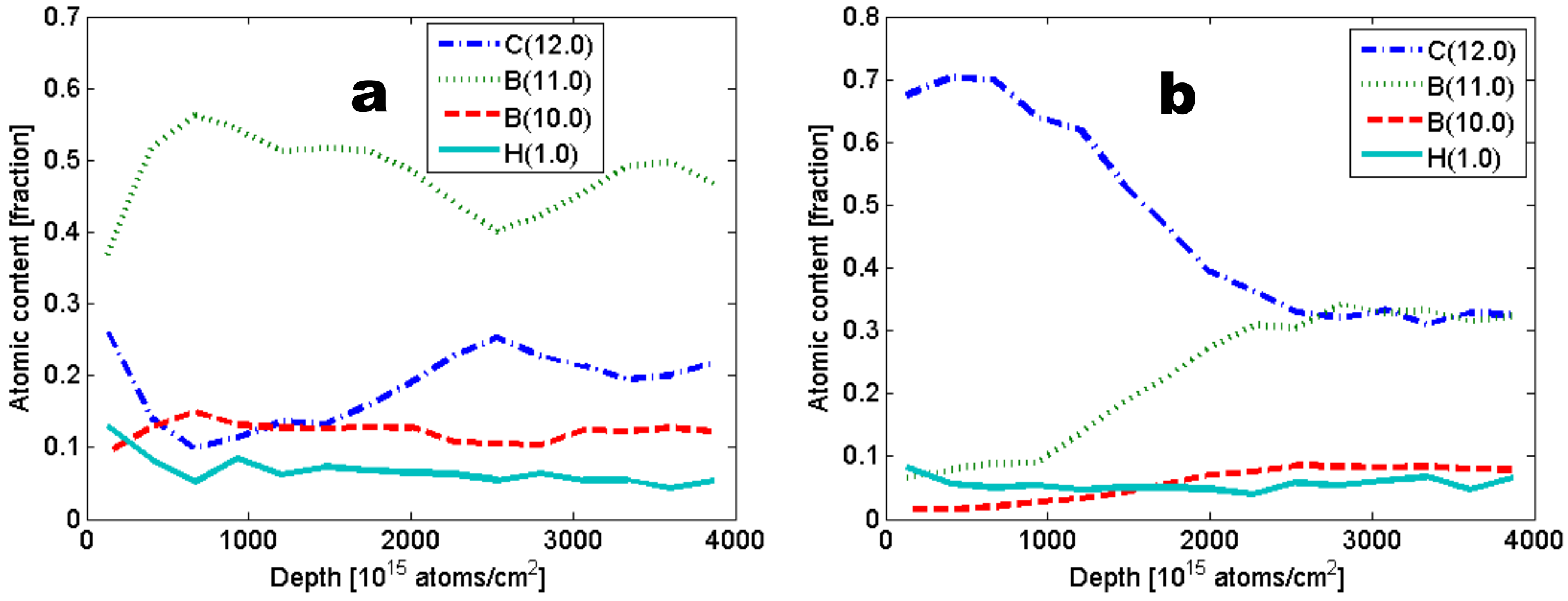

Figure 7. Depth profiles of $^{12}C$, $^{11}B$, $^{10}B$ and H on (a) Tile 3-19, Sample C; (b) Tile 6-19, Sample D

In order to provide a comprehensible overview of the deposited material found on all samples, the depth profiles have been integrated, yielding the number of atoms per unit area for relevant species. Figures 8-10 show results for the amount of respective atoms found in the top 400 nm layer. The absolute numbers are uncertain here, again due to surface roughness introducing an error in the probed layer thickness. Thus, ratios between different elements are the relevant quantities to look at. The diagrams are constructed such that tiles further away from the $MoF_6$ injection point appear further to the right on the x-axis and samples are ordered from A (left) to D (right) for each tile. The samples closest to the injection are, as seen in Figures 1 and 2, 5-28 A, B and 5-14 A, B. Note that Tile 3-19 has very high levels of all studied elements except molybdenum. These abundant deposits were visible as colorful layers on the surface of the tile. Figures 8 a and b show the amount of fluorine and molybdenum respectively, on the surfaces of all analysed samples from ALT-II.

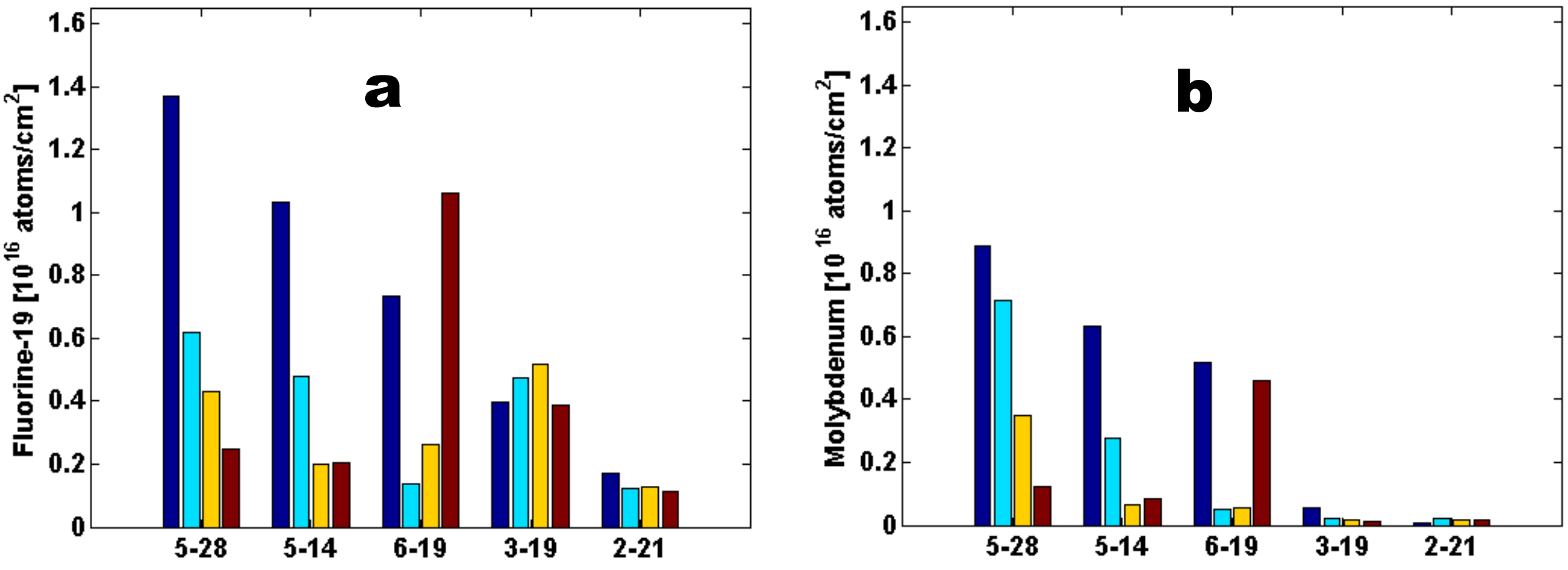

Figure 8. Surface concentrations of (a) fluorine-19 and (b) molybdenum in all samples. The four bars for each tile correspond to Samples A,B,C and D (from left to right).

The greatest content of Mo is on tiles from Blades 5 and 6, located close to the $MoF_6$ injection point. The distribution is not uniform and, especially on Tile 5-28, a substantial amount is detected in the erosion zone, i.e. in the plasma-wetted region. This deposition pattern reflects the mechanism of material migration which was identified earlier in the study of tungsten transport following $WF_6$ puffing [9]. It involves deposition, re-erosion and prompt re-deposition of heavy species. The latter phenomenon of prompt re-deposition was theoretically considered by Naujoks [45], while experiments with high-Z metal puffing allowed for the verification of such a transport mechanism from the plasma-wetted surface to adjacent areas. This statement is supported by the fact that tiles located further from the injection contain much less molybdenum: the measured value is less than $6 \times 10^{14}$ $cm^{-2}$ (i.e at background level) for all samples on Tiles 3-19 and 2-21. The fluorine content also peaks at Tile 5-28, but it is distributed more uniformly on the toroidal limiter than molybdenum. A fluorine distribution of this type is related to the long-range transport of low-Z elements, whose Larmor radius after the ionization in the plasma edge is small making their prompt re-deposition less efficient.

The amounts of nitrogen isotopes: $^{14}N$ and $^{15}N$, detected on the surfaces are given in Figures 9 a and b. In general, there is more $^{14}N$ though it was not used in the marker experiment. As for $^{15}N$, it should be stressed that despite the 90 degree separation of its injection location and the $MoF_6$ inlet, strongly elevated concentrations were found on Tiles 5-28, 5-14 and 6-19 together with Mo. This indicates that in the process of a local deposition of heavy atoms (i.e. Mo), light species are efficiently captured, thus enhancing co-deposition.

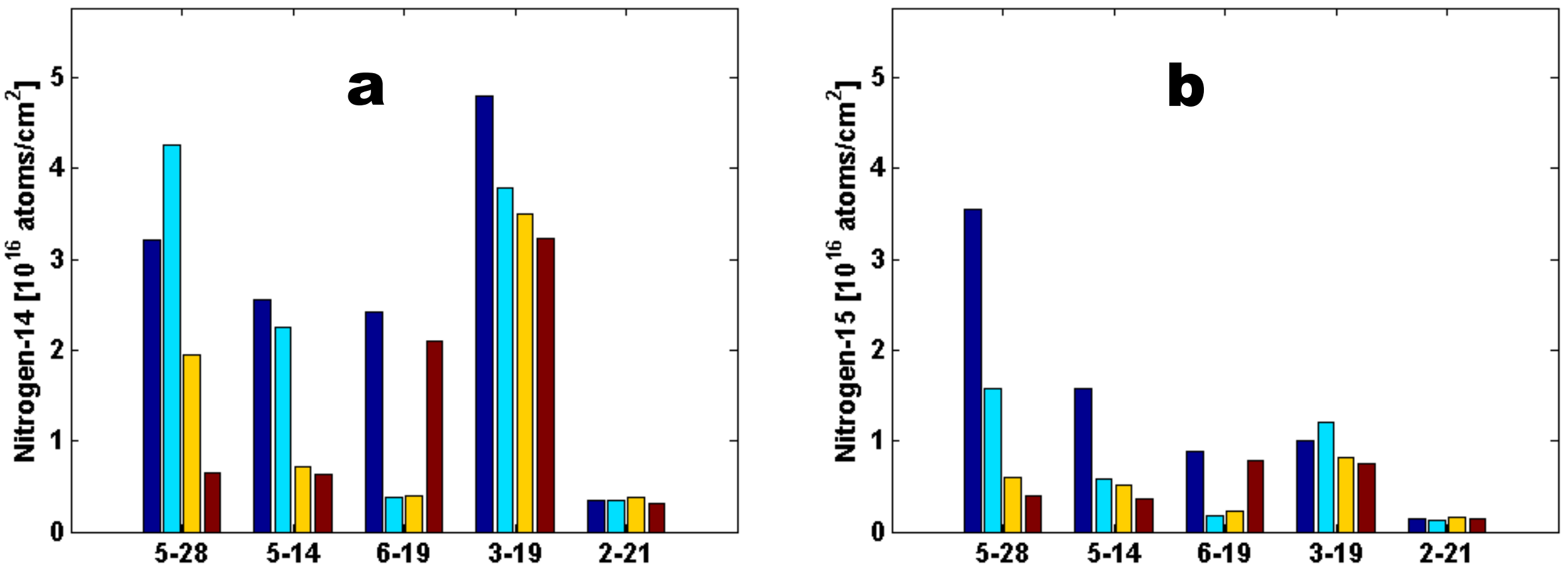

Figure 9. Surface concentration of nitrogen isotopes: (a) nitrogen-14 and (b) nitrogen-15, Samples A,B,C and D from left to right for each tile.

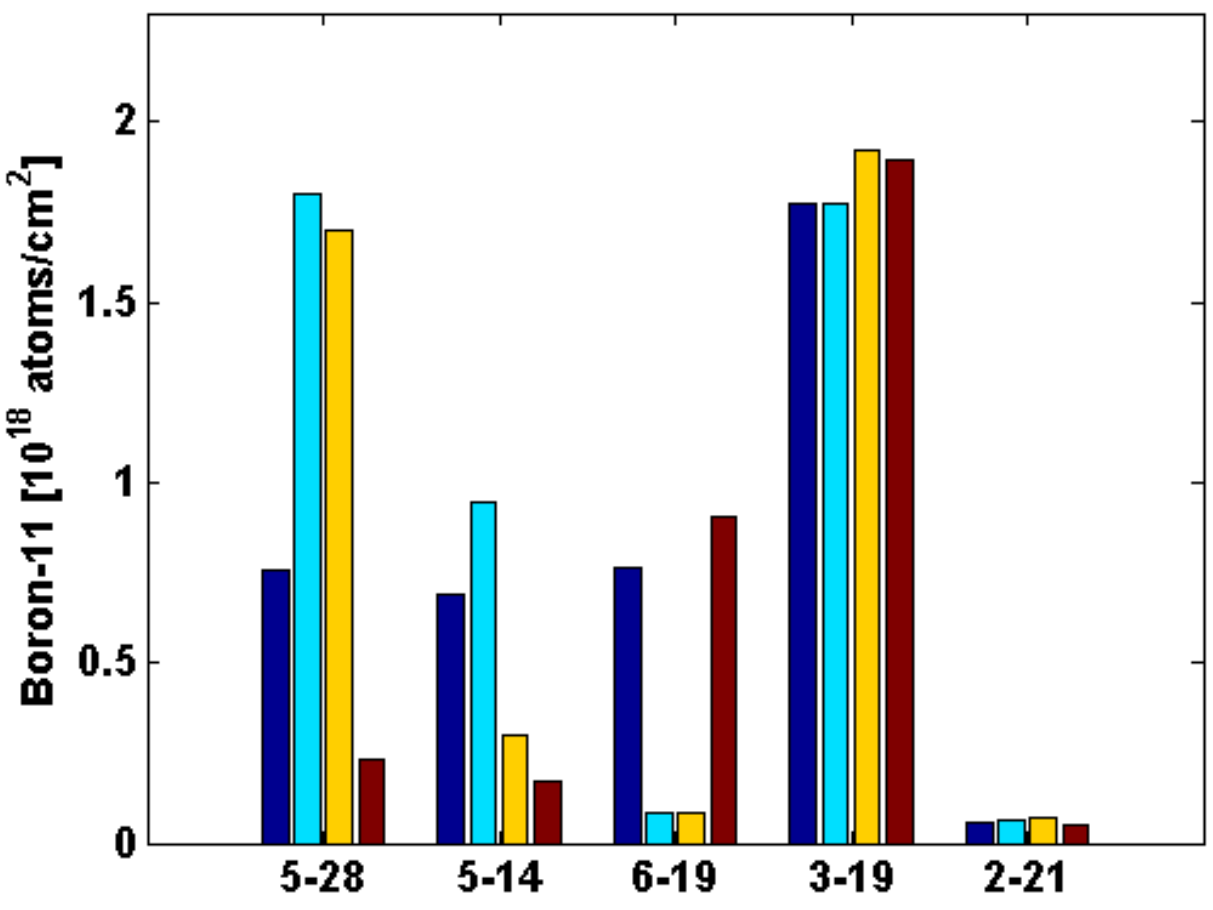

Figure 10. Surface concentration of boron-11, Samples A,B,C and D from left to right for each tile.

Fig. 10 shows the concentration of $^{11}B$. The data clearly proves that the distribution of boron, which is a standard “resident” in TEXTOR [31], differs (especially for tiles from Blade 5) from the distributions of species introduced in the last experiment before the tile retrieval and surface analysis. The boron deposition pattern is rather representative of the entire experimental campaign. From the point of view of this particular work with ToF-HIERDA we stress that despite the problems with surface roughness, the measured

$^{11}B/^{10}B$ atomic concentration ratio is, as expected, very close to the ratio in natural boron with 80.1% of $^{11}B$ and 19.9% of $^{10}B$. The average deviation from this ratio in our measurements is 2.9%.

For all isotopes presented in Figures 8-10, we note that the previously reported erosion zone and deposition zone pattern [5,9,33] is only clearly seen on Tile 6-19, with strongly elevated concentrations on samples A and D.

**4. Concluding remarks**

The work performed is the first one featuring a very broad approach to surface studies of plasma-facing components with ToF-HIERDA. The data clearly indicate a particular usefulness of that method in the determination of material migration mechanisms using tracers, although care has to be taken about surface roughness. The method is especially useful when low-Z isotopes are applied, e.g. $^{15}N$. Most important is a proper separation of adjacent traces in the spectrum because any improper accountancy would be the major source of errors. The ongoing development of a new detection system for ToF-HIERDA [46] will enhance research capabilities and a new set-up with that system will enable analyses of whole tiles without cutting them into smaller specimens. This certainly does not limit the research field to PFCs. On the contrary, it opens and offers opportunities for using tracers in many other fields. For instance, when the mechanism of surface reactions and in-depth migration involving carbon, nitrogen and oxygen is to be determined, one can use $^{13}C$, $^{15}N$ or $^{18}O$. Other important isotopes are $^{10}B$ and $^{6}Li$. One is also given a rare opportunity to determine depth profiles of hydrogen isotopes.

**Acknowledgments**

This work has been carried out within the framework of the EUROfusion Consortium and has received funding from the European Union's Horizon 2020 research and innovation programme under grant agreement number 633053. The views and opinions expressed herein do not necessarily reflect those of the European Commission. The work has been partly funded by the Swedish Research Council (VR) through contract no. 621-2009-4138.

## FIGURE CAPTIONS

**Figure 1.**

(a) Toroidal view inside the TEXTOR vacuum vessel with all limiters (including the test limiter) marked, (b) a magnified image of the $MoF_6$ inlet and (c) a scheme of the ALT-II limiter seen from above, with the location of the $MoF_6$ gas inlet marked by a small circle and the analysed tiles marked with crosses.

**Figure 2.**

Sample positions on ALT tiles: (a) Blade 5, Tile 14: upper corner tile,

(b) Other tiles (including lower corner Tile 5-28, where same scheme was applied). Directions are such that the right edges of tiles here displayed are pointing in the clockwise direction in Fig. 1c.

**Figure 3.**

ToF-HIERDA spectrum from Tile 5-14, Sample A. Intensity scale: number of counts in channel relative to maximum number of counts.

**Figure 4.**

Counts corresponding to $^{14}N$ and $^{15}N$ for (a) Sample 3-19C, with a moderate amount of $^{15}N$ and (b) Sample 5-28A, where $^{15}N$ concentration is comparable to that of $^{14}N$.

**Figure 5.**

Surface profiles on (a) Sample 6-19A (graphite), and (b) Sample 5-14D (CFC). Obtained with a Wyko 9300 optical profilometer. Note that the thin spikes visible especially in (b) are unphysical artefacts from the measurement.

**Figure 6.**

Depth profiles of Mo, $^{19}F$ and $^{15}N$ on (a) Tile 5-28, Sample A and (b) Tile 6-19, Sample D

**Figure 7.**

Depth profiles of $^{12}C$, $^{11}B$, $^{10}B$ and H on (a) Tile 3-19, Sample C; (b) Tile 6-19, Sample D

**Figure 8.**

Surface concentrations of (a) fluorine-19 and (b) molybdenum in all samples. The four bars for each tile correspond to Samples A,B,C and D (from left to right).

**Figure 9.**

Surface concentration of nitrogen isotopes: (a) nitrogen-14 and (b) nitrogen-15, Samples A,B,C and D from left to right for each tile.

**Figure 10**

Surface concentration of boron-11, Samples A,B,C and D from left to right for each tile.

## TABLES

Table 1. List of tiles selected for detailed analyses and justification of the selection.

| **Tile (ALT-II blade number – tile number)** | **Description** |
|---|---|
| 5-28 | Closest tile to $MoF_6$ injection point, visible deposited layers. |
| 5-14 | Close to $MoF_6$ injection point, visible deposited layers. |
| 6-19 | Large molybdenum deposits seen using RBS. |
| 3-19 | Far from $MoF_6$ injection point, but clearly visible deposits. |
| 2-21 | Far from $MoF_6$ injection point, no visible deposits (reference tile). |

Table 2. Surface roughness on six samples quantified by average deviation from the mean surface.

| **Sample** | **Description** | **$R_a$, scan 1 [µm]** | **$R_a$, scan 2 [µm]** | **$R_a$, scan 3 [µm]** |
|---|---|---|---|---|
| 5-14C | CFC tile, erosion zone | 8.31 | 8.40 | 10.50 |
| 5-14D | CFC tile, deposition zone | 5.41 | 6.21 | 6.49 |
| 6-19A | Graphite tile, deposition zone | 3.21 | 3.49 | 3.30 |
| 6-19C | Graphite tile, erosion zone | 1.91 | 2.22 | 2.24 |
| 2-21C | Graphite tile, erosion zone | 2.48 | 2.35 | 2.10 |
| 2-21D | Graphite tile, deposition zone | 1.81 | 1.87 | 1.80 |